\documentclass[nofootinbib,prd,twocolumn,eqsecnum,showpacs,showkeys,preprintnumbers,altaffilletter]{revtex4-1} 
\usepackage{amsmath}
\usepackage{amssymb}
\usepackage{graphicx}
\usepackage{epstopdf}
\usepackage{amsfonts}
\usepackage{color}
\usepackage{bm}
\usepackage{mathrsfs}
\usepackage{url}

\usepackage[normalem]{ulem}
\usepackage{bigints}

\makeatletter

\newcommand*{\rom}[1]{\expandafter\@slowromancap\romannumeral #1@}

\newcommand{\be}{\begin{equation}}
  \newcommand{\ee}{\end{equation}}
\newcommand{\ben}{\begin{eqnarray*}}
  \newcommand{\een}{\end{eqnarray*}}
\newcommand{\bea}{\begin{eqnarray}}
  \newcommand{\eea}{\end{eqnarray}}
\newcommand{\bdm}{\begin{displaymath}}
  \newcommand{\edm}{\end{displaymath}}
\newcommand{\ba}{\begin{align}}
  \newcommand{\ea}{\end{align}}

\makeatother

\begin{document}

\title{Constraints on tachyon inflationary models with an AdS/CFT correspondence}

\author{Zahra Bouabdallaoui$^{1}$}
\email{zahraandto@hotmail.com}
\author{Ahmed Errahmani$^{1}$}
\email{ahmederrahmani1@yahoo.fr}
\author{Mariam Bouhmadi-L\'{o}pez$^{2,3,4,5}$}
\email{mbl@ubi.pt, (on leave of absence from UPV and IKERBASQUE.)}
\author{Taoufik Ouali$^{1}$}
\email{ouali\_ta@yahoo.fr}
\date{\today}
\affiliation{
${}^1${Laboratory of Physics of Matter and Radiation, Mohammed I University, BP 717, Oujda, Morocco\\}
${}^2${Departamento de F\'{i}sica, Universidade da Beira Interior, 6200 Covilh\~{a}, Portugal\\}
${}^3${Centro de Matem\'{a}tica e Aplica\c{c}\~{o}es da Universidade da Beira Interior (CMA-UBI), 6200 Covilh\~{a}, Portugal\\}
${}^4${Department of Theoretical Physics University of the Basque Country UPV/EHU. P.O. Box 644, 48080 Bilbao, Spain\\}
${}^5${IKERBASQUE, Basque Foundation for Science, 48011, Bilbao, Spain}}

\begin{abstract}
In order to study the effect of the anti de Sitter/ conformal field theory correspondence (AdS/CFT) on the primordial inflationary era, we consider a universe filled with a tachyon field in a slow-roll regime. In this context, the background and perturbative parameters characterising the inflationary era are related to the standard one by correction terms. We show a clear agreement between the theoretical prediction and the observational data for the above mentioned model. The main results of our work are illustrated for an exponential potential. We show that, for a suitable conformal anomaly coefficient, AdS/CFT correspondence might  leave its imprints on the spectrum of the gravitational waves amplitude with a tensor to scalar ratio, $r$, of the perturbations compatible with Planck data.
\end{abstract}
\maketitle

\section{Introduction}

It is well known that our universe underwent an inflationary phase at its early epochs (cf. for example ref. \cite{Mukhanov}). Based on the existence of a slowly rolling scalar field regime, with energy-momentum tensor dominated by the contribution of the field potential energy, inflation is to date the most compelling solution to many long-standing problems of the big bang cosmology. It provides a causal interpretation of the origin of the observed anisotropy of the cosmic microwave background (CMB) radiation \cite{Hinshaw}, a mechanism for the production of density perturbations required to seed the formation of structures in the universe \cite{Mukhanov} and also describes the generation of primordial gravitational waves as a vacuum fluctuations of space-time.\\

The measure of the spectral index given by the cosmological data, even with high precision, is not enough to discriminate between the large number of inflationary models which are in a good agreement with this measure. Hence, further analysis of a consistent behaviour of the spectral index versus the tensor to scalar ratio or versus the running of the spectral index and/or that of the tensor to scalar ratio versus the running of the spectral index might help to reduce the number of these inflationary models. The recent Planck data \cite{PlanckColl} quoted to 95\% CL a value of the spectral index, $n_s=0.968\pm 0.006$, an upper limit of the tensor to scalar ration $r<0.11$ and a value of the running of the spectral index, $\alpha_s=dn_s/d\ln k=-0.003\pm 0.007$. One of the main goal of this paper is to show the compatibility of AdS/CFT correspondence with observations by investigating the imprints it might leave in the clumpy universe where inflation is driven by a tachyon field.\\

Tachyon fields play a significant role in realising an early inflation phase. Indeed, a universe dominated by a tachyon field, rolling down  its potential slowly, evolves smoothly from a phase of accelerated expansion to a phase dominated by pressure-less matter \cite{Liv1}; i.e. inflation is realised in a natural way. It has been also considered as a candidate for dark matter and dark energy \cite{Gibbons,Fairbairn,Feinstein,Padmanabhan,Shiu,Roy,Panda,Kim,Sugimoto,Minahan,Cornalba,Benaoum}. Tachyons fields are inherent in brane world cosmology \cite{Papa,Farajollahi}, D-branes inflation \cite{Sen}, multi tachyon fields \cite{Piao:2002}, k-inflation \cite{Armendariz,Garriga}, warm inflationary model \cite{Herrera, Kamali} and recently in LQC \cite{Setare,Jian,Setare1,Kui}. Linear perturbations of tachyon field were discussed in \cite{Garriga,Frolov,Souza,Santillan,Herrera1}. Effective potential for tachyon field in string theory were computed in reference \cite{Gerasimov}. Tachyon field were also considered in Hamilton-Jacobi approach \cite{Aghamohammadi} and in a logamediate inflationary model \cite{Ravanpak}. In ref. \cite{Sen1,Dalianis} have considered an exponential potential for the tachyonic field.\\

The AdS/CFT correspondence is a concrete illustration of the holographic principle \cite{hooft96, susskind95, susskind98, witten98} which indicate a mutual relation between theories with gravity on the bulk and theories without gravity on its boundary. Precisely, the AdS/CFT correspondence is a dual description between a higher (d+1)-dimension Anti de Sitter space time (AdS) and a conformal field theory (CFT) on a lower d-dimensional boundary of the AdS space time. Within the framework of string theory, the AdS/CFT correspondence was conjectured by Maldacena \cite{Maldacena} and it is known actually as a gauge/gravity duality. Therefore, it seems that the AdS/CFT duality is a good framework for improving the constraints on the inflationary parameters such as the gravitational wave amplitude parameter, namely through the tensor to scalar ratio. Using the fiduciary technique, the authors of Ref. \cite{Lidsey} constrained the conformal coefficient anomaly characterising this duality by studying the consistency equation in the AdS/CFT correspondence in a universe filled by a scalar field.\\

In this paper we  study the inflationary parameters of the universe in which its dynamic is driven by a tachyonic field in the context of AdS/CFT correspondence where the tachyon field is considered to be rolling down an exponential potential. This potential has been chosen for simplicity. 
Hence our results are, first, compared to those obtained in the case where a scalar inflaton field \cite{Lidsey} is considered rather than a tachyon field and second to those where a tachyon field is considered but in standard cosmology without considering AdS/CFT correspondence \cite{Nozari}. In this paper we constraint the conformal anomaly coefficient  characterising AdS/CFT correspondence by
taking into account the latest Planck data \cite{PlanckColl}.\\

The outline of the paper is as follows: In section II, we setup the basic framework of our approach such as the modified Friedmann equation by the AdS/CFT correspondence and for a dynamical tachyon field. In section III, we calculate observable quantities such as the amplitude of the scalar perturbation, the scalar spectral index, the amplitude of the tensorial perturbations, the tensor spectral index, the tensor-scalar ratio, the running of the spectral index and the slow roll parameters. In section IV, we illustrate our results by invoking an exponential potential. Finally, we present our conclusions in section V.


\section{ The setup }
As an example of AdS/CFT correspondence we start reviewing briefly the formulation of the holographic dual of the RSII scenario \cite{Lidsey} known as AdS/CFT correspondence (for more details see \cite{Kiritsis,Boer}).
In this setup the Friedmann equation becomes \cite{Kiritsis,Lidsey}
\begin{equation}\label{F1}
H^2=\frac{\hat{m}^2_p}{4c}\Big[1+\epsilon \sqrt{1-\frac{\rho}{\rho_{max}} }\;\Big];
\end{equation}
where $\hat{m}^2_p={m^2_p}/{8 \pi}$ is the reduced 4-dimensional 
Planck mass, $\rho$ is the total energy density of the universe, $\rho_{max}=3\hat{m}^4_p/8c$ and $c=\hat{m}_p^6/8M^6$ is the conformal anomaly coefficient which relates the reduced 4-dimensional Planck mass to the 5-dimensional Planck mass $M$.
The standard form of the Friedmann equation can be recovered at the low-energy limit $\rho\ll\rho_{max}$  for the branch $\epsilon=-1$. 
Henceforth, hereafter only this branch will be considered. \\

Furthermore, we assume that the cosmological dynamics of the universe is driven by a tachyon field $\phi$ with a potential $V(\phi)$. The equation of motion of this tachyon field can be obtained for example from the conservation of the energy momentum tensor and reads \cite{Armendariz, Garriga}
\begin{equation}\label{KLEINGORDON}
   \frac{ \ddot{\phi}}{1-\dot{\phi}}+3H\dot{\phi}+\frac{V'}{V}=0
\end{equation}
where a dot corresponds to a derivative with respect to the cosmic time and a prime means a derivative with respect to the tachyon field $\phi$. The energy density and the pressure of this field are given respectively by  \cite{Sen}
\begin{equation}
   \rho=\frac{V}{\sqrt{1-\dot{\phi}^2}},
\end{equation}
and 
\begin{equation}
   P=-V\sqrt{1-\dot{\phi}^2}.
\end{equation}
 
During the inflationary epoch and assuming a slow-roll expansion, $\dot{\phi}^2 \ll 1 $ and $ \ddot{\phi} \ll 3H\dot{\phi}$, the energy density associated to the tachyon field becomes dominated by the potential i.e. $\rho \simeq V$. Hence the Friedmann equation (\ref{F1}) reduces to
\begin{equation}\label{F2}
H^2=\frac{\hat{m}^2_p}{4c}\Big[1-\sqrt{1-x}\Big]
\end{equation}
where $x=V/V_{max}$ is a dimensionless parameter and $V_{max}={3\hat{m}^4_p}/{8c}$. The equation of motion (\ref{KLEINGORDON}) reduces then to
\begin{equation}
   3H\dot{\phi} \simeq -\frac{V'}{V}=-\frac{x'}{x}.
\end{equation}

\section{Perturbations}
In order to study the cosmological perturbations in the slow roll regime, we consider Eqs. (\ref{F1}) and (\ref{KLEINGORDON}) as the effective four-dimensional equations describing the cosmology of a tachyon field. Scalar perturbations of a Friedmann-Lema\^{i}tre-Robertson-Walker background  in the longitudinal gauge are given by
\begin{equation}\label{PM}
ds^2=-(1+2\Phi)dt^2+a^2(t)(1-2\Psi)\delta_{ij}dx^idx^j,
\end{equation}
where $a(t)$ is the scale factor, $\Phi(t,x)$ and $\Psi(t,x)$ are the scalar perturbations. The spatial curvature perturbation on uniform density hypersurfaces is given by $\zeta=-H\delta\phi/\dot{\phi}$ \cite{Lyth}. Within the slow-roll regime, the amplitude of the perturbation mode crosses the Hubble radius during inflation, the field fluctuations  are given by \cite{Liv1}
\begin{equation}\label{}
<\delta\phi^2>=\frac{H^2}{4\pi^2}.
\end{equation}
The power spectrum of the curvature perturbations is related to the curvature perturbation by 
\begin{equation}\label{}
<A_s^2>=\frac{<\zeta^2>}{V},
\end{equation}
 or equivalently by using the above equations 

\begin{equation}\label{amplitude pertuscalaire}
A^2_s=\frac{1}{4\pi^2}\frac{H^4}{\dot{\phi}^2}\frac{1}{V}.
\end{equation}
Using the slow roll condition, Eq. (\ref{amplitude pertuscalaire}) becomes 
\begin{equation}\label{amplitude pertuscalaireaproximation}
A^2_s=\frac{9 H^6}{4\pi^2}\frac{V}{V'^2}=[A^2_S]_{\textrm{std}}G_{c,r}
\end{equation}
where
\begin{equation}\label{}
[A^2_s]_{\textrm{std}}=\frac{1}{12 \hat{m}^6_p \pi^2}\Big(\frac{V^4}{V'^2}\Big),
\end{equation}
and
\begin{equation}\label{corection}
G_{c,r}=\dfrac{27 \hat{m}^6_p H^6}{V^3 }
\end{equation}
are the amplitude of the scalar perturbation of a tachyonic field in standard cosmology (cf. Ref \cite{Nozari}) and the correction term characterising the effect of AdS/CFT correspondence, respectively. We can notice from Eq. (\ref{F2}) that at the low energy limit, $(V \ll V_{max})$, the correction term reduces to one and the standard expression of the amplitude of the scalar perturbation is recovered. The correction term can be rewritten as a function of the $x$-parameter as 
\begin{equation}\label{parametre3}
G_{c,r}=\Big(\dfrac{2}{1+\sqrt{1-x}}\Big)^3.
\end{equation}

Furthermore, the scalar spectral index $n_s$ is given in term of the scalar perturbation amplitude as $n_s -1=\frac{d\ln A_s^2}{d \ln k}$, where the wave number $k$ at the Hubble crossing is related to the number of e-folds by the relation  $d \ln k = dN$. To first order in the slow-roll parameters and by using Eq. (\ref {amplitude pertuscalaireaproximation}), the scalar spectral index can be written as \cite{Nozari}

\begin{equation}\label{ns}
    n_s\approx1-6\varepsilon +2\eta,
\end{equation}
where the slow-roll parameters $\varepsilon$ and $\eta$ are defined as

\begin{equation}\label{parametre1}
\varepsilon =-\frac{\dot{H}}{H^2} \simeq [\varepsilon]_{\textrm{std}}C_{c,r}, 
\end{equation}
and
\begin{equation}\label{PARAMETRE2}
\eta =\dfrac{1}{3H^2}\Big( \dfrac{V''}{V}-\dfrac{1}{2}\dfrac{V'^2}{V^2}\Big) 
 \simeq  \dfrac{[\eta]_{\textrm{std}}}{G^{\frac{1}{3}}_{c,r}},
\end{equation}
where we have denoted the standard slow-roll parameters as $[\varepsilon]_{\textrm{std}}$ and $[\eta]_{\textrm{std}}$ (cf. Ref. \cite{Nozari}), of the tachyonic field respectively as 
\begin{equation}
[\varepsilon]_{\textrm{std}}=\dfrac{ \hat{m}^2_p }{2} \dfrac{V'^2}{V^3},
\end{equation}
and 

\begin{equation}
[\eta]_{\textrm{std}}=\hat{m}^2_p\Big(\dfrac{V''}{V^2}-\dfrac{1 }{2}\dfrac{V'^2}{V^3}\Big).
\end{equation}

The correction term $C_{c,r}$ is related to $G_{c,r}$ by 

\begin{equation}\label{corectionC2}
   C_{c,r}=\dfrac{1}{G^{\frac{2}{3}}_{c,r}\sqrt{1-x}}=\dfrac{(1+\sqrt{1-x})^2}{4\sqrt{1-x}}.
\end{equation}

Using Eqs. (\ref{ns}), (\ref{parametre1}) and (\ref{PARAMETRE2}), the scalar spectral index can be rewritten as a function of the correction term  $G_{c,r} $ and the standard slow roll parameters as: 
\begin{equation}\label{nsgafe5}
  n_s = 1-\dfrac{ 2}{G^{\frac{1}{3}}_{c,r}}  
	\Big[ \frac{3}{G^{\frac{1}{3}}_{c,r}\sqrt{1-x}}[\varepsilon]_{std}-[\eta]_{std} \Big],
\end{equation}
or equivalently in term of the dimensionless parameter $x$ as
\begin{equation}\label{nsgafe}
  n_s = 1- \frac{8c}{3\hat{m}^2_p}\Big(1+\sqrt{1-x} \Big) \Big[ \Big(\frac{3+5\sqrt{1-x}}{4\sqrt{1-x}}  \Big)\frac{x'^2}{x^3}-\frac{x''}{x^2} \Big].
\end{equation}
We can notice from Eq. (\ref{F2}) that at the low energy limit ($x\ll 1$) the correction terms reduce to one. Also
the standard expressions of the scalar spectral index (Eqs. (\ref{ns}) and (\ref{nsgafe5})) as well as those of the slow roll parameters (Eqs. (\ref{parametre1}) and (\ref{PARAMETRE2})) are recovered. \\

On the other hand, the generation of the tensor perturbations during inflation would produce a spectrum of the gravitational waves with an amplitude given by \cite{Nozari}
\begin{equation}\label{amplitperturbationtensoriel}
A^2_T=\frac{4}{25 \pi \hat{m}^2_p}H^2,
\end{equation}
which can be written in term of the standard tachyonic inflationary scenario as:
\begin{equation}\label{amplitperturbationtensoriel2}
A^2_T=\dfrac{4V}{75 \hat{m}^4_p \pi}G^{\frac{1}{3}}_{c,r}=[A^2_T]_{std} G^{\frac{1}{3}}_{c,r},
\end{equation}
where  $[A^2_T]_{std}$  is the amplitude of the tensor perturbation of the tachyonic field in standard cosmology \cite{Nozari} and $G_{c,r}$ is given by Eqs. (\ref{corection}) and (\ref{parametre3}). We can show that at the low energy limit the standard expression of the amplitude of the  tensor perturbation is recovred. \\

The  spectral index related to the tensor perturbation is given by $n_T=\frac{d\ln A_T^2}{d \ln k }$ and using Eq. (\ref{amplitperturbationtensoriel}), we obtain
\begin{equation}\label{}
n_T=-2\varepsilon,
\end{equation}
which can be written in term of the standard slow roll parameter and the correction to standard general relativity as: 
\begin{equation}\label{nT}
n_T=-2 [\varepsilon]_{\textrm{std}}  C_{c,r}.
\end{equation}

The tensor-scalar ratio, $r$, is defined as:
\begin{equation}\label{rts}
r=\dfrac{A^2_T}{A^2_s},
\end{equation}
which becomes using Eqs. (\ref{amplitude pertuscalaire}) and (\ref{amplitperturbationtensoriel})
\begin{equation}\label{rapport tensor scalaire}
r = [r]_{\textrm{std}}\Big(\dfrac{V}{3 \hat{m}^2_p H^2}\Big)^2 
\simeq   [r]_{std}G_{c,r}^{-2/3},
\end{equation}
where $[r]_{\textrm{std}}=\dfrac{16 \hat{m}^2_p }{25\pi}\dfrac{V'^2}{V^3}$
is the standard tensor-scalar ratio produced by a tachyonic field  \cite{Nozari}.\\

The appearance of the correction terms in the above equations (Eqs. (\ref{amplitude pertuscalaireaproximation}), (\ref{parametre1}), (\ref{PARAMETRE2}), (\ref{nsgafe5}), (\ref{amplitperturbationtensoriel2}), (\ref{nT}), and (\ref{rapport tensor scalaire}))  implies that AdS/CFT correspondence  might modify considerably the predicted value of the inflationary parameters, as we will be showing in the figures below.\\

In order to constraint the range of the conformal anomaly coefficient $c$ and the tensor-scalar ratio $r$, we deduce 
from Eqs. (\ref{F2}), (\ref{amplitperturbationtensoriel}) and (\ref{rts}) a relationship between them and the parameter $x$ as
\begin{equation}  \label{epsilon5}
x= \tilde{c} r(2-\tilde{c}r),
\end{equation}
where 
\begin{equation}\label{COEFICIENT}
\tilde{c}= 25 \pi c A^2_s.
\end{equation}
To recover the standard cosmology, the conformal anomaly coefficient is bounded by a maximal value, $c_{max}$, obtained at the singular point where $x=1$ ($V_{hc}=V_{max}$) in which the Hubble parameter, Eq. (\ref{F2}), is finite while its first derivative is infinite ($\ddot{a} \longrightarrow -\infty$) i.e. a sudden singularity is met. Indeed, the condition $x \ll 1$ requires an upper bound on the conformal anomaly coefficient $c$ such that

\begin{equation}  \label{PARANOMALI}
c \ll c_{max}=7.24\times 10^7.
\end{equation}
This maximal value is obtained by using the most recent Planck data \cite{PlanckColl} where $A^2_s=2.20\times 10^{-9}$, for a tensor-scalar ratio $r$ equal to $0.08$ and by equating $x$ to unity in Eq. (\ref{epsilon5}).  \\

\begin{figure}[hbtp]
\centering
\includegraphics[scale=0.4]{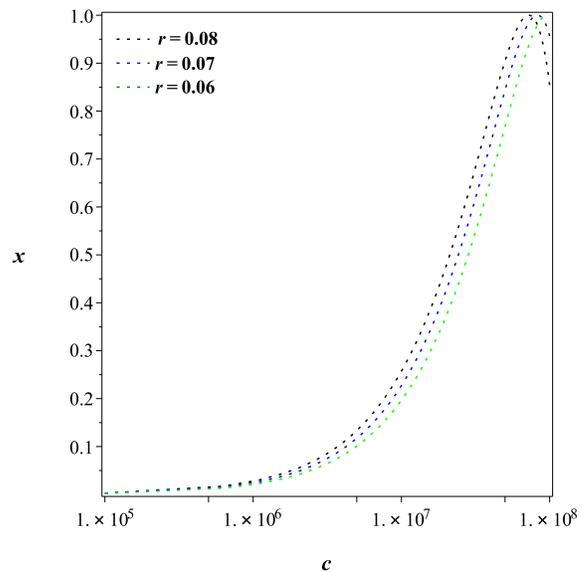} 
\caption{ Evolution of the dimensionless parameter $x$  versus the  conformal anomaly coefficient $c$  for different  values of the tensor-scalar ratio $r$ and for $A^2_s=2.20\times 10^{-9}$.\label{x_c}}
\end{figure}

\begin{figure}[hbtp]
\centering
\includegraphics[scale=0.4]{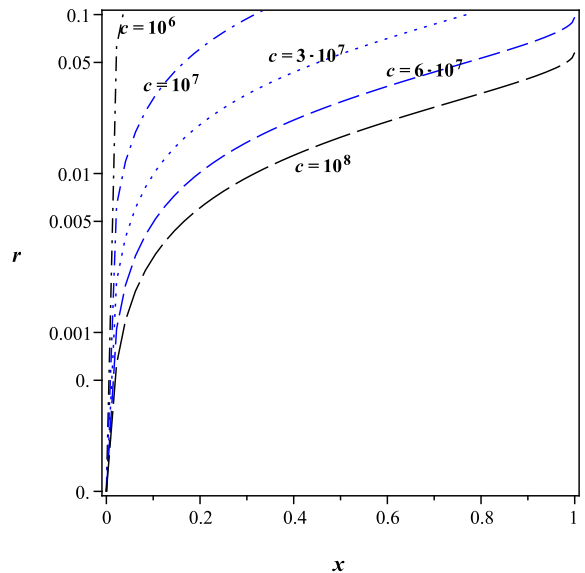}
\caption{ Evolution of the tensor-scalar ratio $r$ versus the dimensionless parameter $x$  for different values of the  conformal anomaly coefficient $c$.}\label{x_r}
\end{figure}

Fig.\ref{x_c} shows the variation of the dimensionless parameter $x$ versus the conformal anomaly coefficient $c$ for different values of the tensor-scalar ratio. We notice that for $c< 10^{6}$, the condition $x\ll 1$ is always fulfilled. In this range, the AdS/CFT correspondence has no effect on the standard cosmology dynamic. This justify the choice of the values of the c-parameter in Fig.\ref{x_r} in which we plot the tensor-scalar ratio versus the dimensionless parameter $x$. We notice that the tensor-scalar ratio  of the curves with the value $c<10^{7}$ is ruled out by Planck data \cite{PlanckColl} for an appropriate $x$-parameter. We conclude, from these figures, that AdS/CFT correspondence can leave its imprints on the spectrum of the  gravitational waves for $c>10^{7}$.

\begin{figure}[hbtp]
\centering
\includegraphics[scale=0.4]{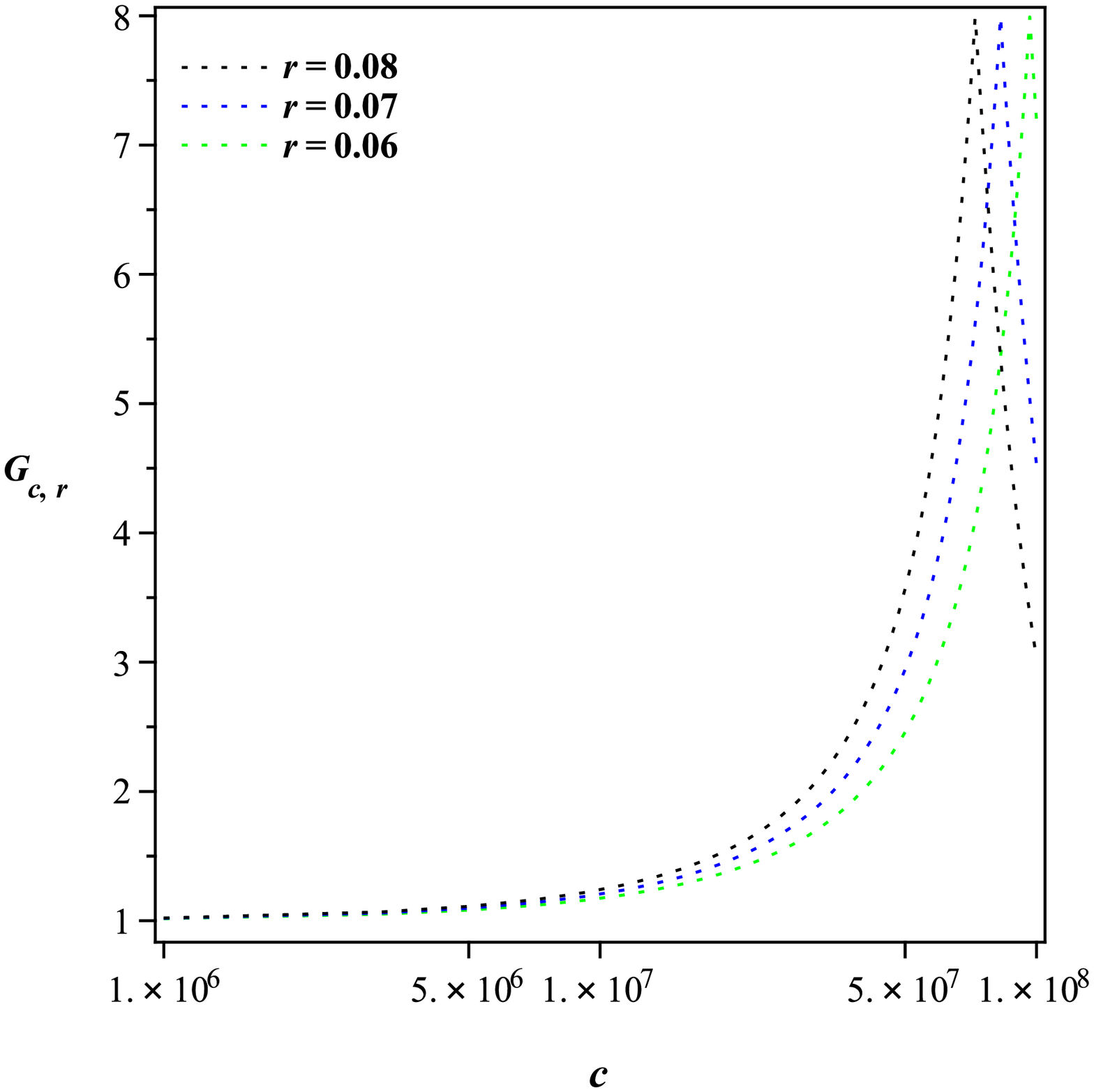} 
\caption{Evolution of the  correction term $G_{r,c}$   versus the conformal anomaly coefficient $c$ for different values of the tensor-scalar ratio $r$.}\label{G_c}
\end{figure}

\begin{figure}[hbtp]
\centering
\includegraphics[scale=0.4]{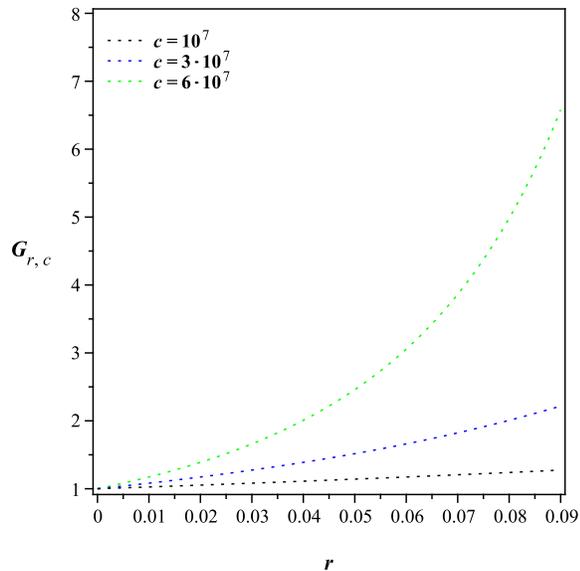} 
\caption{  Evolution of the  correction term $G_{r,c}$   versus the tensor-scalar ratio $r$ for different values of the conformal anomaly coefficient $c$.}\label{G_r}
\end{figure}
Fig.\ref{G_c} shows the variation of the correction term $G_{c,r}$ appearing in the amplitude of the scalar perturbation Eq. (\ref{amplitude pertuscalaireaproximation}) and in the tensor perturbation Eq. (\ref{amplitperturbationtensoriel2}) versus the conformal anomaly coefficient $c$ for different values of the tensor-scalar ratio. We notice again that for $c< 10^{7}$, the correction term is of the order of unity which means that the results are not affected by AdS/CFT correspondence. This justify the choice of the values of the c-parameter in Fig.\ref{G_r} for the correction term, $G_{c,r}$, versus the tensor-scalar ratio for different values of the conformal anomaly coefficient $c$. We notice that the value $c=10^{7}$ (the blue curve) confirm the fact that the correction term equal to 1 for all values of the tensor to scalar ratio. While the value $c>10^{7}$ gives an appreciable correction term for a tensor-scalar ratio in the range of the observed data. We conclude from these figures that again AdS/CFT correspondence can leave some finger prints on the amplitude of the cosmological perturbations for $c>10^{7}$. The maximum in  Figs. \ref{G_c} shows the maximal value of the conformal anomaly coefficient corresponding to $x=1$ and to a correction term equal to 8 (e.g. for $r=0.08$ Fig. \ref{G_c} shows $c_{max}\approx 6.$ $10^{7}$). 

\begin{figure}[hbtp]
\centering
\includegraphics[scale=0.4]{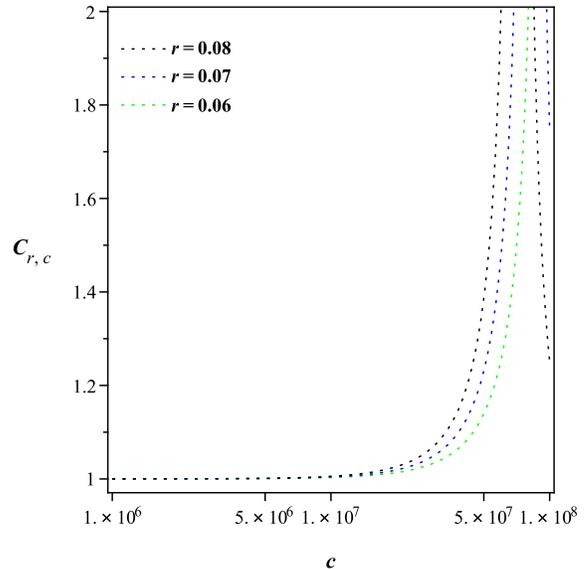} 
\caption{  Evolution of the  correction term $C_{r,c}$   versus the conformal anomaly coefficient $c$ for different values of the tensor-scalar ratio $r$.}\label{C_c}
\end{figure}
\begin{figure}[hbtp]
\centering
\includegraphics[scale=0.4]{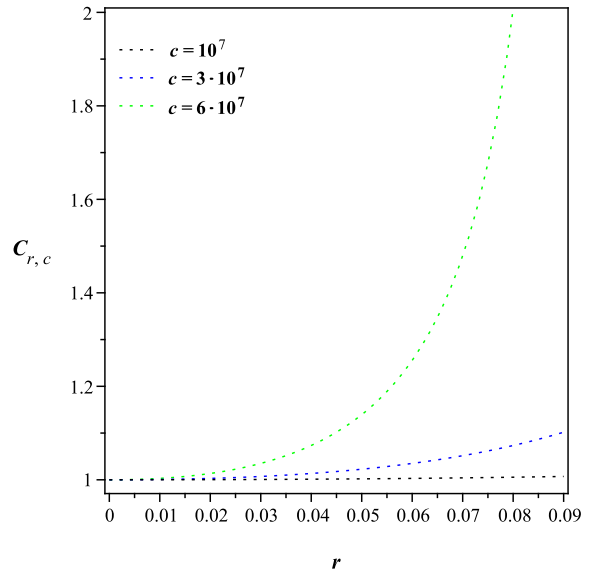} 
\caption{  Evolution of the  correction term $C_{r,c}$   versus the tensor-scalar ratio $r$ for different values of the conformal anomaly coefficient $c$.}\label{C_r}
\end{figure}

Fig.\ref{C_c} shows the variation of the correction term $C_{c,r}$ appearing in the slow roll parameter Eq. (\ref{parametre1}) versus the conformal anomaly coefficient $c$ for different values of the tensor-scalar ratio. We notice that for $c < 10^{7}$,  the correction term is of the order of unity which means that the results are not affected by AdS/CFT correspondence. This again justify the choice of the values of the c-parameter in Fig.\ref{C_r} for the correction term, $C_{c,r}$, versus the tensor-scalar ratio. We notice that the value $c=10^{7}$ (the blue curve) gives a correction term equal to one while it begins to be different from one for $c>10^{7}$  and $r>0.02$. We conclude from these figures that AdS/CFT correspondence can once more imprint some net feature on the amplitude of the  gravitational waves for values of the c-parameter bigger than $10^{7}$. The maximum in  Figs. \ref{C_c} corresponds to the maximal value of the conformal anomaly coefficient corresponding to $x=1$ where the expression of the correction term Eq. (\ref{corectionC2}) is not defined. We can notice also that these maximums shift in the sense of the increasing values of the conformal anomaly coefficient and of a decreasing values of the tensor to scalar ratio as shown in Fig.\ref{C_r}.\\

We summarise this section by concluding that the effect of AdS/CFT correspondence on the background and perturbative parameters characterising the inflationary era can be observed for the allowed conformal anomaly coefficient in the range $10^7<c<10^8$.

\section{exponential potential}

The exponential potential within the inflationary scenario is motivated
by supergravity and string theory \cite{Goncha,Stew,Dval}. 
Such a potential has a maximum value at the tachyon field $\phi=0$ corresponding in some bosonic string  to a tension of an unstable D-brane while its local minimum, $V=0$ at $\phi\rightarrow\infty$, corresponds to a closed bosonic string.
In this section, we consider such an exponential potential satisfying these properties and given by \cite{Liv1,Kamali,Jian,Setare1,Fairb}

\begin{equation}  \label{epsilon}
V=V_0\exp( -\alpha \phi)
\end{equation}
where $\alpha$ is a parameter related to the mass of the tachyon field \cite{Liv1} and  $V_0$ corresponds to the maximum value of the potential.  \\

\begin{figure}[hbtp]
\centering
\includegraphics[scale=0.4]{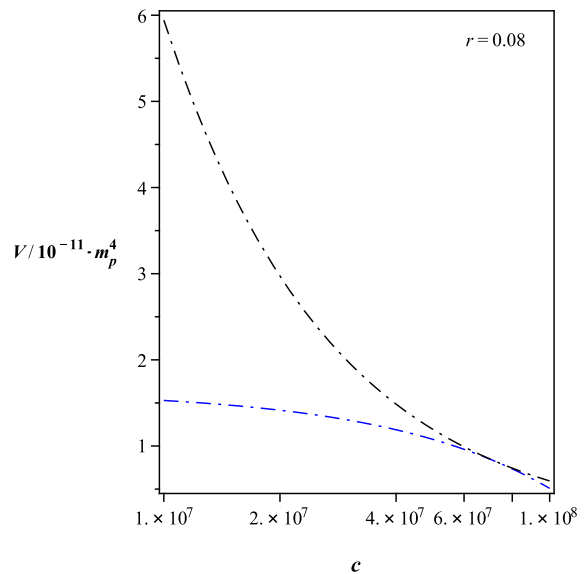} 
\caption{Evolution of the potentials $V_{hc}$ (blue curve) and $V_{max}$ (black curve) versus the  conformal anomaly coefficient $c$ for $r=0.08$.}\label{Vhc1}
\end{figure}

In Fig. \ref{Vhc1}, we show the variation of the potential at the horizon crossing  $V_{hc}$ obtained from Eq. (\ref{epsilon5}) and the variation of the potential $V_{max}$ equal to $3\hat{m}_p^4/8c$ versus the conformal anomaly coefficient $c$ for $r=0.08$. We notice that the maximum value of the conformal anomaly coefficient characterising the bound of the AdS/CFT duality  appears around the numerical value obtained in Eq. (\ref{PARANOMALI}).\\

Furthermore, the amount of inflation is measured by the e-folding number $N$ defined by \cite{Nozari} 
\begin{eqnarray}\label{N-efold}
N&=&\int_{t_{hc}}^{t_f}Hdt \nonumber \\
 &\simeq &- \int_{V _{hc}}^{V _ f}3H^2 \frac{V}{V'^2}dV
\end{eqnarray}
where the subscript "${hc}$" denotes the value of a given quantity at the horizon crossing during inflation
and "$f$" its value when the universe exits the inflationary phase. By integrating Eq. (\ref{N-efold}), we obtain

\begin{equation}  \label{nbredefolding}
N=\frac{3\hat{m}^2_p}{2c\alpha^2} \Big (\ln (\frac{1+\sqrt{1-x_{hc}}}{1+\sqrt{1-x_f}})+\sqrt{1-x_f}-\sqrt{1-x_{hc}}  \Big ), 
\end{equation}
where $x$ is defined in Eq. (\ref{F2}).

\begin{figure}[hbtp]
\centering
\includegraphics[scale=0.4]{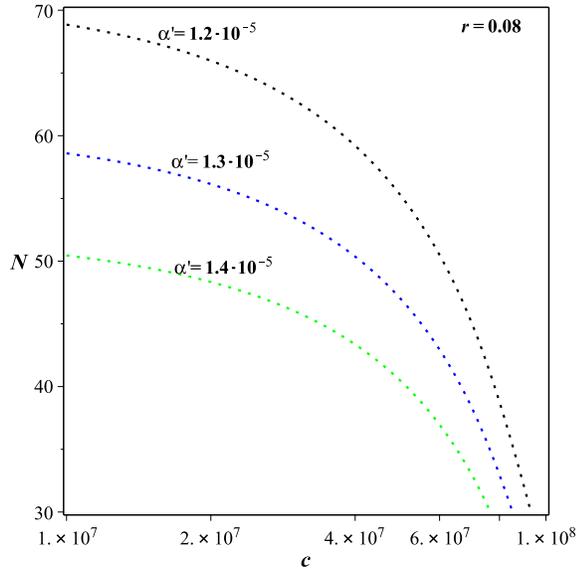} 
\caption{  Evolution of the  e-folding number $N$  versus the conformal anomaly coefficient $c$ for different values of the $\alpha$-parameter and for the tensor-scalar ratio $r=0.08$. We set $\alpha'=\alpha/\hat{m}_p$.}\label{N_c1}
\end{figure}

\begin{figure}[hbtp]
\centering
\includegraphics[scale=0.4]{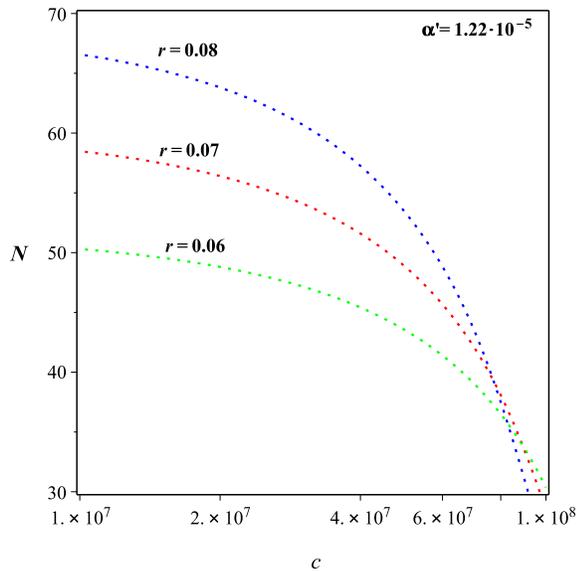} 
\caption{  Evolution of the  e-folding number $N$   versus the conformal anomaly coefficient $c$ for different values of the tensor-scalar ratio $r$ and for $\alpha'=\alpha/\hat{m}_p=1.22\times 10^{-5}$.}\label{N_c2}
\end{figure}

From Eq. (\ref{parametre1}) the slow roll parameter $\varepsilon$ becomes equal to
\begin{equation}\label{epsilon_F}
\varepsilon=\dfrac{c\alpha^2}{3\hat{m}^2_p}\dfrac{x}{\sqrt{1-x}(1-\sqrt{1-x})^2},
\end{equation}
from which we can obtain the value of the quantity $x_f$ at the end of inflation i.e. at $\varepsilon=1$. We can show that at the low energy limit we obtain $x_f={4c\alpha^2}/{3\hat{m}^2_p}$, which is exactly the standard expression of the potential obtained in Ref \cite{samiPRD66}, namely $V_f=\alpha^2\hat{m}_p^2/2$.\\
The main parameters of our model are shown in table I for $r=0.08$. For the $\alpha$-parameter value and for the e-folds N ($50$ or $60$), we obtain the conformal anomaly coefficient from Fig \ref{N_c1} from which we can deduce $V_{max}$. Also from Eqs. (\ref{epsilon5}) and (\ref{epsilon_F}), we deduce $V_{hc}$ and $V_{f}$, respectively. As we can notice from Fig \ref{N_c1}, the e-folds $N$never reaches $70$ for an appropriate conformal anomaly coefficient and for a tensor to scalar ratio bounded by observational data \cite{PlanckColl}.\\
\begin{table} \label{1}
\begin{tabular} { |c |c |c |c |c |c |c |c || }
\hline  \hline
$\alpha/\hat{m}_p$ & $N$ & $c/10^7$ & $V_{max}/\hat{m}_p^4$   & $V_{hc}/\hat{m}^4_p$ &  $V_{f}/\hat{m}_p^4$ \\
\hline\hline

\small $1.23\times 10^{-5}$ & $50$ & \small $5.57$ &\small $6.732\times 10^{-9}$ &\small $6.374\times 10^{-9}$ & \small $7.564\times 10^{-11}$ \\

 &\small$60$ & \small $2.9$ & $1.293\times 10^{-8}$ & \small $8.286\times 10^{-9}$ & \small $7.564 \times 10^{-11}$    \\\hline

\small $  1.3\times 10^{-5}$ & $50$ & \small $4.1$ &\small $9.14\times 10^{-9}$ &\small $7.42\times 10^{-9}$ & \small $8.45\times 10^{-11}$ \\

 &\small $60$ & \small $0.35$ & $1.07\times 10^{-7}$ & \small $1.01\times 10^{-8}$ & \small $8.44\times 10^{-11}$   \\\hline 

\end{tabular}
\caption{The main parameters of the model for $r=0.08$.}
\end{table}
Figs. \ref{N_c1} and \ref{N_c2} show suitable values of the conformal anomaly coefficient, namely around $10^{7}$, for which the e-folding number $50<N<70$ for different values of the $\alpha$-parameter for the tensor to scalar ratio $r=0.08$ (Fig. \ref{N_c1}) and for different values of the tensor-scalar ratio for the $\alpha$-parameter equal to $1.5\times 10^{-5}\hat{m}_p$ (Fig. \ref{N_c2}). \\

In order to study the behaviour of the tensor-scalar ratio versus the scalar spectral index, we  
calculate the scalar perturbation from Eq. (\ref{amplitude pertuscalaire}) at the crossing horizon
\begin{equation}  \label{AMPLITUDEDEPERTURBATION}
A^2_s=\frac{3\hat{m}^2_p}{32\pi^2 \alpha^2 c^2} \frac{(1-\sqrt{1-x_{hc}})^3}{x_{hc}},          
\end{equation}
the tensor-scalar ratio from Eq. (\ref{rapport tensor scalaire}), i.e.,
\begin{equation}  \label{rapport}
r= \frac{32\pi c \alpha^2  }{75 \hat{m}^2_p} \frac{x_{hc}}{(1-\sqrt{1-x_{hc}})^2},
\end{equation}
and the scalar spectral index from Eq. (\ref{nsgafe}) reads
\begin{equation}  \label{rapport23}
n_s=1-\frac{ 2 c \alpha^2 }{3 \hat{m}^2_p} \Big[ \frac{3+ \sqrt{1-x_{hc}}}{\sqrt{1-x_{hc}}(1-\sqrt{1-x_{hc}})}\Big].
\end{equation}
\begin{figure}[hbtp]
\centering
\includegraphics[scale=0.4]{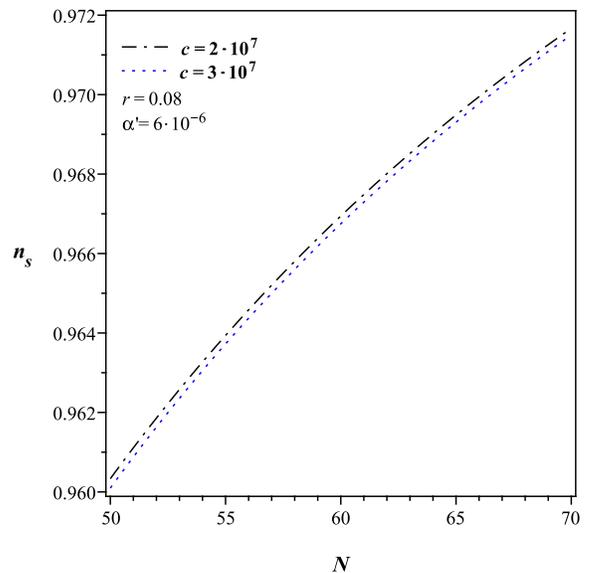} 
\caption{Evolution of the scalar spectral index $n_s$ versus the number of e-folds $N$ for different values of $c$, for $r=0.08$ and $\alpha=6\times 10^{-6}\hat{m}_p$.}\label{ns_N_alpha}
\end{figure} 

Fig.\ref{ns_N_alpha} shows the variation of the scalar spectral index $n_s$ versus the number of e-folds for different values of the conformal anomaly coefficient for $\alpha=0.6\times 10^{-5}\hat{m}_p$ and for $r=0.08$. We notice that for the e-folding number $50<N<70$, the spectral index reaches the observed value for the conformal anomaly coefficient used in the plot. \\

Also in order to take into account the variation of the scalar spectral index $n_s$, we consider its running given by $\alpha_s=\frac{dns}{dlnk}$. From Eqs. (\ref{nbredefolding}) and (\ref{rapport23}), we obtain the running spectral index as
\begin{equation}  \label{rapport}
\alpha_s= \frac{4c^2 \alpha^4}{9 \hat{m}^4_p}\Big (\frac{3(x_{hc}-\sqrt{1-x_{hc}})- (1-x_{hc})(4+\sqrt{1-x_{hc}})}{(2x_{hc}+ (1-x_{hc})^\frac{3}{2})(2-x_{hc})-2} \Big )      .
\end{equation}

\begin{figure}[hbtp]
\centering
\includegraphics[scale=0.4]{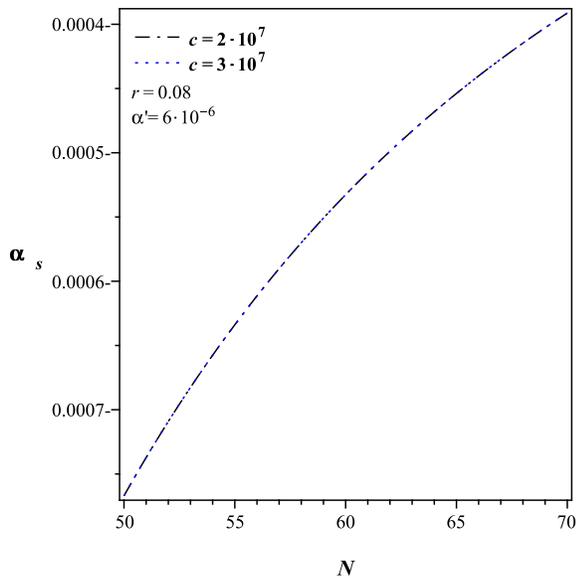} 
\caption{Evolution of the running spectral index $\alpha_s$ versus  the number of e-folds $N$, for different values of $c$, for $r=0.08$ and $\alpha=6 \times 10^{-6}\hat{m}_p$ .}\label{alphas_N}
\end{figure} 

\begin{table}[h!]\label{table2}
\begin{tabular}{|c|c|c|c|c|}
\hline \hline
      & $N=50$     & $N=60$ & $N=70$ &Planck TT,TE,EE+lowP \\ \hline
$n_s$      & $0.96$    & $0.966$ & $0.971$ &$\small 0.968\pm 0.006$ \\ \hline
$\alpha_s$ & $-0.00076$ & $-0.00053$ &$-0.00039$  &$-0.003\pm 0.007$        \\  \hline 
\end{tabular}
\caption{Comparison of the observed and the predicted (for $N=50$, $N=60$ and $N=70$) values of the perturbative parameters of the inflationary era, $n_s$ and $\alpha_s$ for $\alpha=6\times 10^{-6}\hat{m}_p$, $c=2\times 10^7$ and $r=0.08$.} 
\end{table}
In table II, we compare the observed and the predicted values of the perturbative parameters of the inflationary era. These results are obtained from Figs. \ref{ns_N_alpha} and \ref{alphas_N} and show an agreement between the observed and the predicted inflationary parameters.

\begin{figure}[h!]
\centering
\includegraphics[scale=0.4]{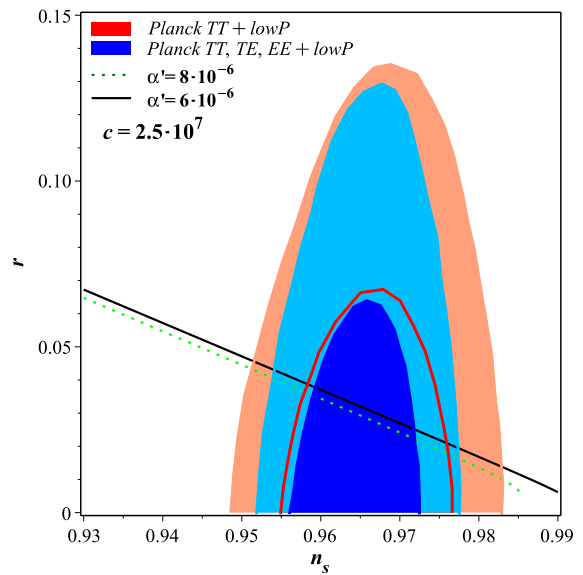} 
\caption{Plot of the parameter $r$ as a function of the scalar spectral index $n_s$ for two values of the
$\alpha$-parameter: $\alpha= 0.8\times 10^{-5}\hat{m}_p$ (green dashed line) and $\alpha= 0.6\times 10^{-5}\hat{m}_p$ (black solid line) and for the conformal anomaly coefficient $c= 2.5\times 10^7$. The marginalized joint 68\% and 95\% confidence level contours $(n_s,r)$ using Planck TT + low P, Planck TT, TE, EE + low P data release are shown \cite{PlanckColl}.}\label{r_ns_cf}
\end{figure}
\begin{figure}[h!]
\centering
\includegraphics[scale=0.4]{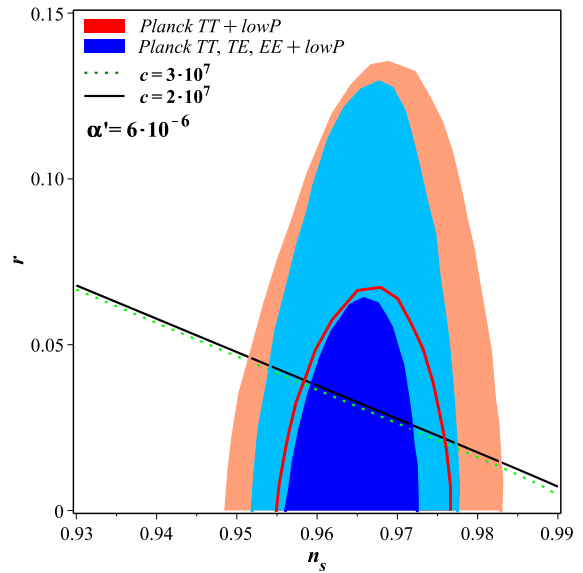} 
\caption{Plot of the parameter $r$ as a function of the scalar spectral index $n_s$ for two values of the conformal anomaly coefficient: $c=3\times 10^7$ (green dashed line) and $c=2\times 10^7$ (black solid line) and for the $\alpha$-parameter $\alpha= 0.6\times 10^{-5}\hat{m}_p$. The marginalized joint 68\% and 95\% confidence level contours $(n_s,r)$ using Planck TT + low P, Planck TT, TE, EE + low P data release are shown \cite{PlanckColl}.}\label{r_ns_cv}
\end{figure} 
\begin{figure}[h!]
\centering
\includegraphics[scale=0.4]{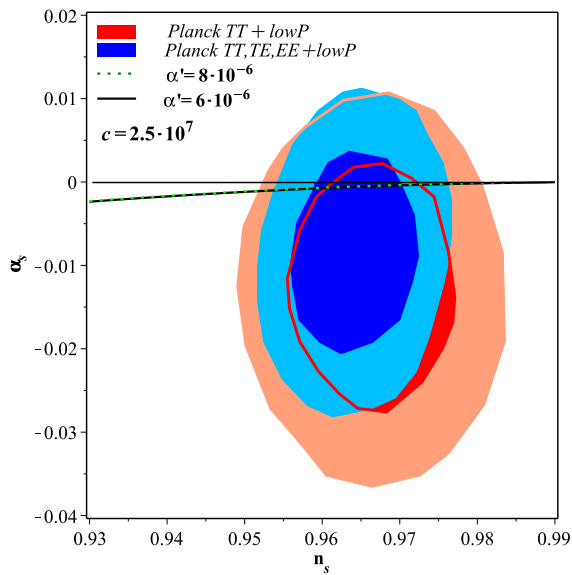} 
\caption{Plot of the running spectral index $\alpha_s$ as a function of the scalar spectrum index $n_s$, for different two values of the $\alpha$-parameter $\alpha= 0.8\times 10^{-5}\hat{m}_p$ (green dashed line) and $\alpha= 0.6\times 10^{-5}\hat{m}_p$ (black solid line) and for the conformal anomaly coefficient $c=2.5\times 10^{7}$. The marginalized joint 68\% and 95\% confidence level contours $(n_s, dns/d \ln k)$ using Planck TT + low P, Planck TT, TE, EE + low P data release are shown \cite{PlanckColl}.}\label{alfas_ns_alfav}
\end{figure} 
\begin{figure}[h!]
\centering
\includegraphics[scale=0.4]{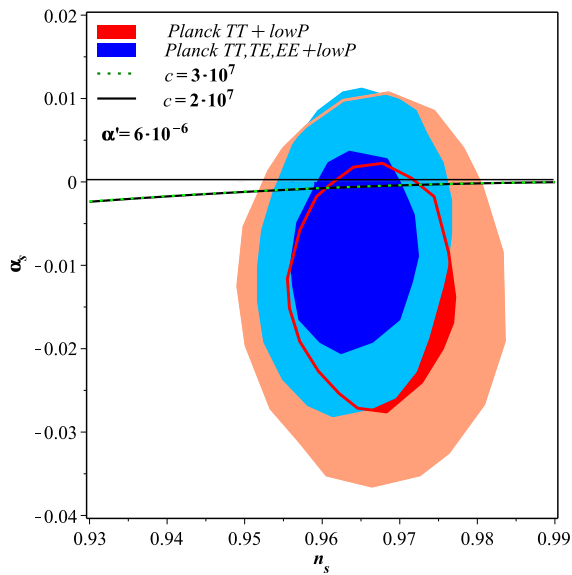} 
\caption{Plot of the running spectral index $\alpha_s$ as a function of the scalar spectrum index $n_s$, for two values of the conformal anomaly coefficient $c=3\times 10^7$  (green dashed line) and $c=2\times 10^7$ (black solid line) and for the $\alpha$-parameter $\alpha= 0.6\times 10^{-5}\hat{m}_p$. The marginalized joint 68\% and 95\% confidence level contours $(n_s, dns/d\ln k)$ using Planck TT + low P, Planck TT, TE, EE + low P data
release are shown \cite{PlanckColl}.}\label{alfas_ns_cv}
\end{figure} 
Fig \ref{r_ns_cf} shows the variation of the tensor-scalar ratio $r$ as a function of the scalar spectrum index $n_s$, for the $\alpha$-parameter equal to $0.8\times 10^{-5}\hat{m}_p$ (green dashed line) and $0.6\times 10^{-5}\hat{m}_p$ (black solid line) for the conformal anomaly coefficient $c=2.5\times 10^7$. Similarly, Fig \ref{r_ns_cv} shows the variation of the tensor-scalar ratio $r$ as a function of the scalar spectrum index $n_s$, for the conformal anomaly coefficient $c=3\times 10^7$ (green dashed line) and $c=2 \times 10^7$ (black solid line) for the $\alpha$-parameter equal to $0.6\times 10^{-5}\hat{m}_p$. \\

In Figs \ref{r_ns_cf} and \ref{r_ns_cv}, we represent our theoretical predicted results based on the Planck data \cite{PlanckColl},  by plotting the evolution of the tensor to scalar ratio versus the scalar spectral index (straight line). As we notice, the predicted parameters  of the AdS/CFT correspondence as well as of the exponential potential lie in the core of the data i.e. in the contour at the 95\% C.L.. We can conclude that the description of the perturbative parameters of the inflationary era of a universe filled by a tachyon field in the context of AdS/CFT correspondence is fully consistent with the recent observational data for the $\alpha$-parameter of the order of $10^{-5}$. \\

In the same way, we plot the evolution of the running spectral index versus the scalar spectral index in Figs. \ref{alfas_ns_alfav} and \ref{alfas_ns_cv} in order to represent our theoretical predicted results based on Planck data \cite{PlanckColl}. One of the plot is done for the $\alpha$-parameter equal to $0.8\times 10^{-5}\hat{m}_p$ (green dashed line) and $0.6\times 10^{-5}\hat{m}_p$ (black solid line) for the conformal anomaly coefficient $c=2.5\times 10^7$ (Fig. \ref{alfas_ns_alfav}). A corresponding situation (Fig. \ref{alfas_ns_cv}) is realized for the conformal anomaly coefficient  $c=3 \times 10^7$ (green dashed line) and $c=2\times 10^7$ (black solid line) for the $\alpha$-parameter equal to $0.6\times 10^{-5}\hat{m}_p$. As we notice, the predicted parameters  of AdS/CFT correspondence as well as of the exponential potential lie in the core of the data i.e. in the contour at the 95\% C.L.. We can conclude that the description of the inflationary parameters of a universe filled by a tachyon field in the context of AdS/CFT correspondence is consistent with the recent observational data for the $\alpha$-parameter of the order $10^{-5}$.   We can also show that a conformal anomaly coefficient higher than $c_{max}$ gives a positive values of the running spectral index $\alpha_s$.\\ 

Furthermore, the $\alpha$-parameter used above lies in the range of Ref. \cite{Liv1} and of the one estimated by the authors in Ref. \cite{samiPRD66}. We should  also notice that the conformal anomaly coefficient obtained above is of the order of the values used in \cite{Lidsey}.\\
\section{Conclusion}
In this paper, we have studied the inflationary scenario for a universe filled with a tachyon  field in the context of AdS/CFT correspondence. Such a tachyon field drives the primordial inflationary era. 
The effect of AdS/CFT correspondence on the perturbative parameters of the inflationary era is characterised by the conformal anomaly coefficient $c$. \\

We have considered that the cosmological dynamics of the tachyon field is responsible for the primordial inflationary era and have assumed an exponential potential characterised by a free parameter $\alpha$. \\

We have found that in order to reproduce the standard cosmology the values of the conformal anomaly coefficient $c$ should satisfy an upper limit, Eq. (\ref{PARANOMALI}). Furthermore, we have shown in Fig. \ref{x_c} and \ref{x_r} the range of the conformal anomaly coefficient and of the tensor-scalar ratio for which the imprints of AdS/CFT correspondence appears clearly at the perturbative level.\\

We have shown that the background and the perturbative  parameters of the inflationary scenario are equal to the standard one times some corrections term which at the low energy limit tends to one. The correction terms are drawn in Figs \ref{G_c} and \ref{C_c} as a function of the conformal anomaly coefficient for different values of the tensor to scalar ratio and as a function of the tensor to scalar ratio for different values of the conformal anomaly coefficient in Figs \ref{G_r} and \ref{C_r}.  These plots show and confirm the fact that the correction terms tend to one for an appropriate conformal anomaly coefficient and tensor to scalar ratio.\\

We have compared our theoretical prediction with observational data \cite{PlanckColl} for $\alpha\approx 10^{-5}\hat{m}_p$ and for the conformal anomaly coefficient $c\approx 10^7$ by plotting the evolution of different inflationary parameters. We have shown that the predicted inflationary parameters lie in the core of the confidence level contours and hence they are consistent with the observational data for the selected range of the conformal anomaly coefficient (Figs \ref{r_ns_cf}-\ref{alfas_ns_cv}).\\ 

We conclude also that the AdS/CFT correspondence  may describe the inflationary era in a universe filled with a tachyon field and predicts the appropriate inflationary parameters with respect to the observational data in the slow roll regime for an allowed range of the conformal anomaly coefficient ($10^7<c<10^8$).\\ 

In a forthcoming paper, we shall study the consistency equation of the inflationary scenario in this approach and look for a possible departure from standard cosmology within the AdS/CFT correspondence.
\acknowledgments
The work of A.E. and T.O. is supported by CNRST, through the fellowship URAC 07/214410. A.E and T.O would like to thank professor D. Khattach for his technical help.
The work of M.B.L. is supported by the Portuguese Agency 
``Funda\c{c}\~ao para a Ci\^encia e Tecnologia" 
through an Investigador FCT Research contract, with reference IF/01442/2013/
CP1196/CT0001. She also wishes to acknowledge the partial support from the
Basque government Grant No. IT592-13 (Spain) and FONDOS FEDER under grant
FIS2014-57956-P (Spanish governmen).

\end{document}